\documentclass[sigconf]{acmart}

\newcommand{\tabincell}[2]{\begin{tabular}{@{}#1@{}}#2\end{tabular}}
\usepackage{balance}

\def\BibTeX{{\rm B\kern-.05em{\sc i\kern-.025em b}\kern-.08emT\kern-.1667em\lower.7ex\hbox{E}\kern-.125emX}}
\usepackage{subfigure} 
\usepackage{hyperref}

\copyrightyear{2019}
\acmYear{2019}
\acmConference[MM '19]{Proceedings of the 27th ACM International Conference on
Multimedia}{October 21--25, 2019}{Nice, France}
\acmBooktitle{Proceedings of the 27th ACM International Conference on Multimedia (MM '19),
October 21--25, 2019, Nice, France}
\acmPrice{15.00}
\acmDOI{10.1145/3343031.3351059}
\acmISBN{978-1-4503-6889-6/19/10}

\begin{document}
\fancyhead{}


\title{Understanding the Teaching Styles by an Attention based Multi-task Cross-media Dimensional modelling}

\author{Suping Zhou}
\affiliation{%
  \institution{Department of Computer Science and Technology}
  \institution{Tsinghua University}
}
\email{1874504489@qq.com}

\author{Jia Jia}
\authornote{corresponding author.}

\affiliation{%
  \institution{Tsinghua University}
  \institution{Beijing National Research Center for Information Science and Technology}
}
\email{jjia@mail.tsinghua.edu.cn}

\author{Yufeng Yin}
\affiliation{%
  \institution{Tsinghua University}
}
\email{yinyf15@mails.tsinghua.edu.cn}

\author{Xiang Li}
\affiliation{%
  \institution{Tsinghua University}
}
\email{li-xiang16@mails.tsinghua.edu.cn}

\author{Yang Yao}
\affiliation{%
  \institution{Tsinghua University}
}
\email{yanghaiqie@outlook.com}

\author{Ying Zhang}
\affiliation{%
  \institution{Tsinghua University}
}
\email{zhangy0018@163.com}

\author{Zeyang Ye}
\affiliation{%
  \institution{Tsinghua University}
}
\email{yezeyang17@gmail.com}

\author{Kehua Lei}
\affiliation{%
  \institution{Tsinghua University}
}
\email{492808177@qq.com}

\author{Yan Huang}
\affiliation{%
  \institution{TAL Education Group}
}
\email{galehuang@100tal.com}

\author{Jialie Shen}
\affiliation{%
  \institution{Queen's University of Belfast}
}
\email{j.shen@qub.ac.uk}

%
\begin{abstract}
Teaching style plays an influential role in helping students to achieve academic success. In this paper, we explore a new problem of effectively understanding teachers' teaching styles. Specifically, we study 1) how to quantitatively characterize various teachers' teaching styles for various teachers and 2) how to model the subtle relationship between cross-media teaching related data (speech, facial expressions and body motions, content et al.) and teaching styles. Using the adjectives selected from more than 10,000 feedback questionnaires provided by an educational enterprise, a novel concept called Teaching Style Semantic Space (TSSS) is developed based on the pleasure-arousal dimensional theory to describe teaching styles quantitatively and comprehensively. Then a multi-task deep learning based model, Attention-based Multi-path Multi-task Deep Neural Network (AMMDNN), is proposed to accurately and robustly capture the internal correlations between cross-media features and TSSS. Based on the benchmark dataset, we further develop a comprehensive data set including 4,541 full-annotated cross-modality teaching classes. Our experimental results demonstrate that the proposed AMMDNN outperforms (+0.0842 in terms of the concordance correlation coefficient (CCC) on average) baseline methods. To further demonstrate the advantages of the proposed TSSS and our model, several interesting case studies are carried out, such as teaching styles comparison among different teachers and courses, and leveraging the proposed method for teaching quality analysis.
\end{abstract}

\maketitle

\section{Introduction}

Teaching effectiveness has a significant influence on student's learning performance \cite{atkins2002effective}. In real practice, teachers assist students to achieve learning objectives mainly through verbal and nonverbal behaviours with multiple channels \cite{gorham1988relationship}. 
Generally, the teaching behavior and the teaching beliefs matching to it can be defined as \emph{teaching style}\cite{heimlich2002teaching}\cite{heimlich1994developing}, which can be conveyed by teachers' speech, body motion, as well as teaching contents during the class.
Consequently, this inspires us about potential of leveraging cross-media data to model teachers' verbal and nonverbal behaviours and gain comprehensive understanding of teaching styles.

Teaching is complex process and thus how to develop advanced technique to characterize teaching styles based on multi-modal learning is not a trivial task. Several related attempts have been witnessed. Zhou \textit{et al.} \cite{zhou2018inferring} Propose a Multi-path Generative Neural Network which considers both acoustic and textual features for emotion inferring. Chen \textit{et al.} \cite{chen2018twitter} Propose a novel scheme for Twitter sentiment analysis with textual information and extra attention to emojis.
Focus on style analysis, Kwon \textit{et al.} \cite{kwon2003emotion} uses Gaussian SVM to conduct a four-class speaking styles classification of three stressed styles (angry, Lombard and loud) and a neutral style. Mohammadi \textit{et al.}
\cite{mohammadi2011humans} applies prosodic features and personality assessments to distinguish between professional and non-professional speaking styles. 
However, works about teaching styles inferring are limited, and most of them are single modal analysis \cite{kwon2003emotion}\cite{smiljanic2009speaking}.

In this paper, we aim to leverage cross-media information about teaching classes including acoustic, visual and textual data to achieve effective teaching style understanding. Towards this end, we address two important questions:
1) how to quantitatively characterize various teachers' teaching styles for various teachers, 2) how to model the subtle relationship between cross-media teaching related data and teaching styles.
In order to solve the above challenges, first，a two-dimensional Teaching Style Semantic Space (TSSS) is built to describe teaching styles quantitatively and comprehensively based on the pleasure-arousal dimensional theory proposed by \cite{mehrabian1980basic}.
Then based on Thulac \cite{sun2016thulac}, the most often used 41 teaching style adjectives are selected from more than 10,000 feedback questionnaires provided by Tomorrow Advancing Life (TAL) Education Group \footnote{http://en.100tal.com/ a leading education and technology enterprise in China}, and manually labeled them on the TSSS.
Also an Attention based Multi-path Multi-task Deep Neural Network (AMMDNN) is proposed to accurately and robustly capture the internal correlations between cross-media features and TSSS, which consists of pleasure-arousal values, and adjective words.
Employing the benchmark dataset we build with 4,541 cross-media teaching classes data collected from TAL, 
an extensive range of tests have been designed to evaluate the mapping effects between cross-media features and coordinate values in the TSSS. 
The results indicate that the proposed AMMDNN model outperforms (+0.0842 in terms of CCC on average) several alternative baselines. Meanwhile, by linking the two-dimensional coordinates with teaching style adjectives, we further show that our method can help to describe the teaching styles more reasonably and vividly. 
Finally, we also carry out several interesting case studies including teaching styles comparison among different teachers and courses, and leveraging the proposed method for teaching quality analysis. Given the importance of teaching style, this study can serve as a springboard for further scholarly exploration.

Our contributions can be summarized as follows:
\begin{itemize}
  \item We explore a new problem of effectively understanding teachers' teaching styles based on multi-modal learning. To our best knowledge, this is the first attempt to study the problem of understanding teachers' teaching styles automatically. Moreover, some interesting case studies are carried to show the effectiveness of our proposed method for understanding teaching styles, such as teaching styles comparison among different teachers and courses, teaching quality analysis.
   \item To gain a comprehensive and flexible teaching styles description, we adopt a dimensional method, a universal Teaching Style Semantic Space (TSSS) to describe teaching styles quantitatively. It is a two-dimensional space containing 41 words that people often use to describe teaching styles. Based on the TSSS, we can not only quantitatively analyze the teachers' teaching styles but also describe the teaching styles more reasonably and vividly.
    \item We propose an Attention based Multi-path Multi-task Deep Neural Network (AMMDNN), to implement the task of mapping cross-media features of teachers in class to the TSSS. Specifically, first we use a multi-path solution to avoid high dimensional input with limit training data, then we adopt attention mechanism to merge the high-level representations of multi-modal features which can better capture the cross-modality correlations in element samples. Furthermore, we regard predicting pleasure and arousal values as two related tasks to further improve the performance. 
\end{itemize}
\vspace*{-0.5\baselineskip}
 The rest of paper is organized as follows. Section 2 lists related works. Section 3 formulates the problem. Section 4 presents the methodologies. Section 5 introduces the dataset, experiment results and case studies. Section 6 is the conclusion.
\vspace*{-0.5\baselineskip}

\section{Related Work}
\textbf{Analysis on personal styles.} Many efforts have been made on personal styles analysis in different application domain~\cite{ShenSCT09}. 
In \cite{wurtzel1971evaluation}, Wurtzel \textit{et al.} state that acting styles for television and cinema tend to be controlled and naturalistic, while stage acting style is expansive and somewhat exaggerated. 
 Smiljanic \textit{et al.} \cite{smiljanic2009speaking} develop a method to elicit three different speaking styles: reduced, citation, and hyper-articulated.
 Using isolated words recorded at an 8kHz sampling rate in various speaking styles, Kwon \textit{et al.} \cite{kwon2003emotion} define three stressed styles (angry, Lombard and loud) and one neutral style, and use Gaussian SVM to conduct four-class speaking style classification. 
Eyben \textit{et al.} \cite{eyben2007wearable} utilize SVM to classify different dance styles like Waltz, Viennese Waltz, Tango, Quick Step, Foxtrot, Rumba, Cha Cha, Samba and so on.
However, these studies mainly focus the categorical styles analysis, which may limit the diversity of personal styles. 

\textbf{Cross-media dimensional modelling.} 
There are two main types of cross-media modelling strategies: categorical and dimensional ones. For categorical modelling,
Chen \textit{et al.} \cite{chen2018twitter} propose a novel scheme for Twitter sentiment analysis with textual information and extra attention on emojis. Zhou \textit{et al.} \cite{zhou2018inferring} propose a Multi-path Generative Neural Network which considers both acoustic and textual features for emotion inferring. 
Since categorical modelling may limit the diversity of personal styles, plenty of previous works based on dimensional modelling have been done in recent years. 
Thammasan \textit{et al.} \cite{thammasan2017multimodal} present a multi-modal study of fusion of EEG and musical features in recognition of arousal and pleasure values for music.
Chen \textit{et al.} \cite{chen2017multimodal} apply a multi-task learning strategy for multiple kinds of para-linguistic information with shared representations. 

Nevertheless, works about how to build a semantic space for teaching styles analysis are limited, and most of them are based on single modality analysis. Therefore, how to map cross-media information in the teaching classes to the teaching styles accurately is still a problem.

\begin{figure*}[t]
\centering
\includegraphics[width=0.95\textwidth]{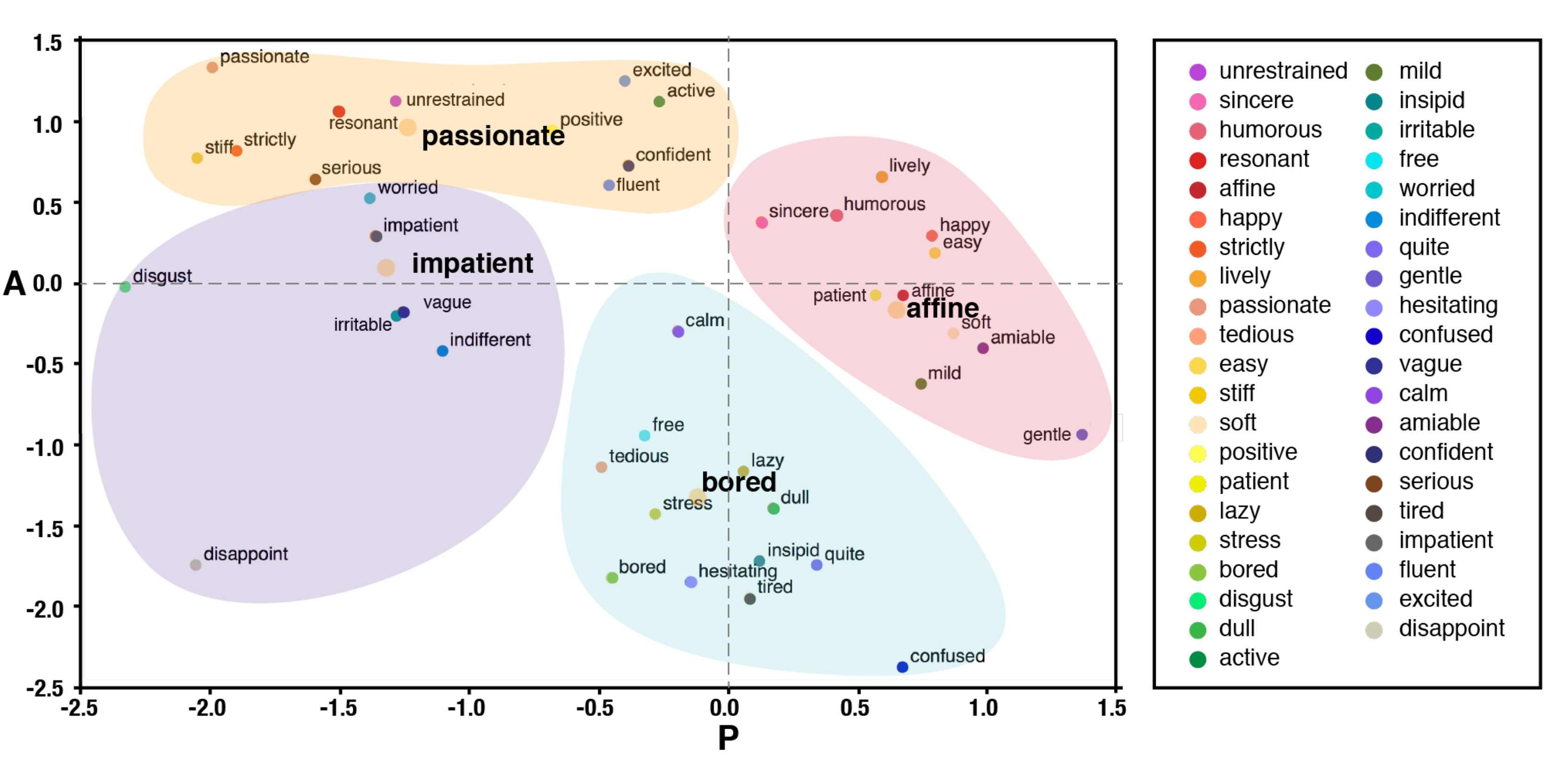}
\vspace*{-0.5\baselineskip}
\caption{The Teaching Style Semantic Space}
\label{fig:tsss}
\end{figure*}

\section{Problem Formulation}

Given a set of utterances $V$, for each utterance $v \in V$, we denote $v=\{x^a,x^t,x^v\}$. $x^a$ represents the acoustic features of each utterance, which is a $n_a$ dimensional vector. $x^t$ represents the textual features of each utterance, which is a $n_t$ dimensional vector.$x^v$ represents the visual features of each utterance, which is a $n_v$ dimensional vector. In addition, $X^a$ is defined as a $|V|*n_a$ feature matrix with each element $x^a_{ij}$ denoting the $j$th acoustic feature of $v_i$. The definition of $X^t$ and $X^v$ is similar to $X^a$.

\emph{Definition} \textbf{The teaching style semantic space} - We adopt a two-dimensional space (pleasure and arousal), denoted as $D(p, a)$. The horizontal axis represents coordinate value for pleasure, while the vertical axis represents coordinate value for arousal.

\emph{Problem} \textbf{Learning task} - Given utterances set $V$, we aim to infer the coordinate value in the Teaching Style Semantic Space for every utterance $v \in V$:
\vspace*{-0.5\baselineskip}
\begin{equation}
    f:(V,X^a,X^t,X^v) \Rightarrow D(p, a)
    \vspace*{-0.5\baselineskip}
\end{equation}

\section{Methodology}

In this study, teaching styles understanding can be formulated with three subtasks: 1) Building a Teaching Style Semantic Space (TSSS) to describe teaching styles systematically and quantitatively; 2) Utilizing different regression models to build the mapping from cross-media features to the TSSS; 3) Linking the two-dimensional coordinates with teaching style adjectives.
\vspace*{-0.5\baselineskip}
\subsection{Building the Teaching Style Semantic Space}

Inspired by \cite{mehrabian1996pleasure}, a two-dimensional semantic space is proposed to describe various teachers' teaching styles quantitatively. Specifically, the semantic space has two dimensions. The horizontal axis refers to pleasure value indicating how pleasant or unpleasant teachers feel, while the vertical axis refers to arousal value indicating how energized or soporific teachers feel. However, two questions need to be addressed for building the TSSS: 1) What adjectives should we use in the semantic space? 2) How can we map the teaching style adjectives to the semantic space?

\begin{figure}[t]
\centering
\includegraphics[width=0.45\textwidth]{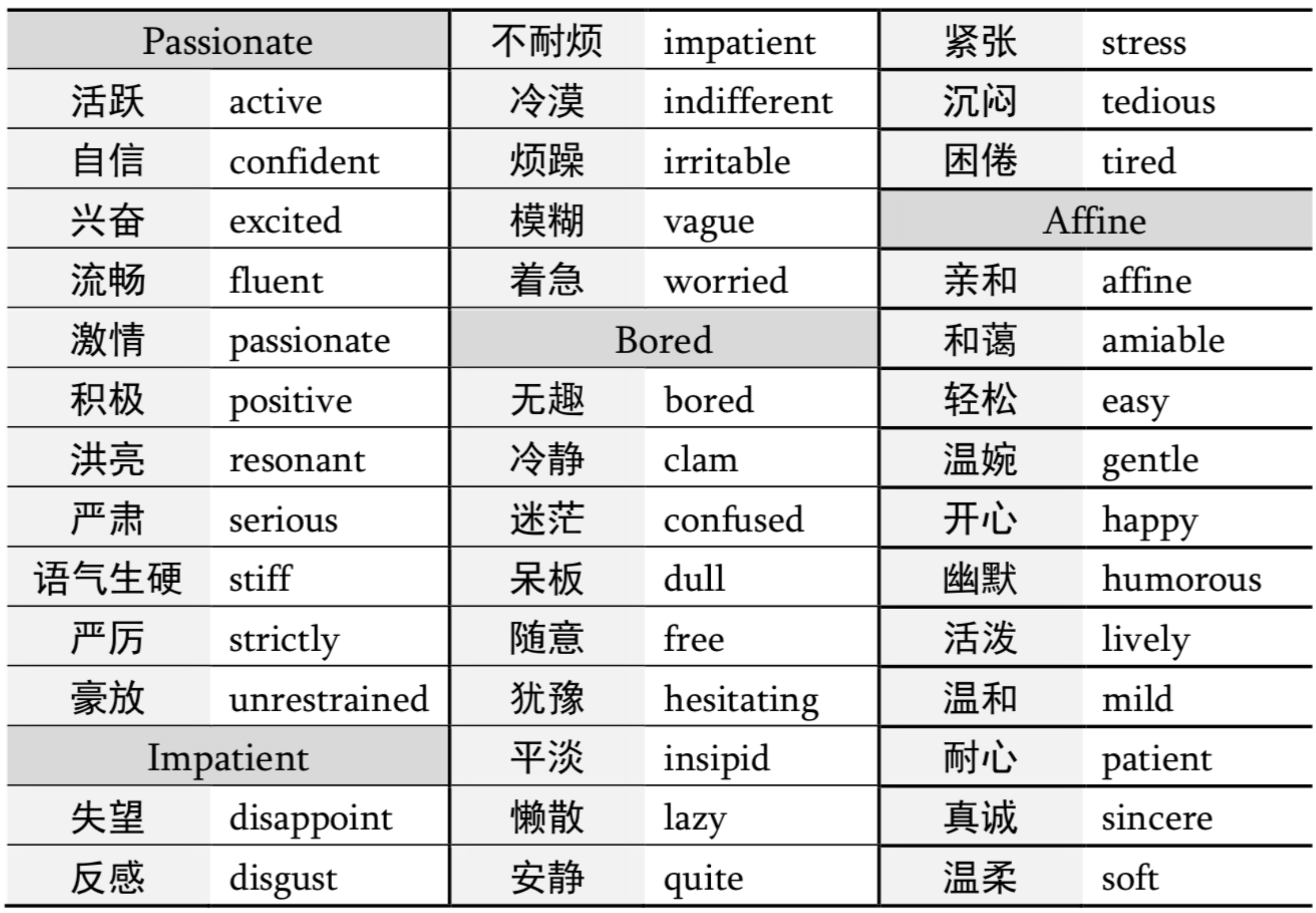}
\vspace*{-0.5\baselineskip}
\caption{41 Teaching Style Adjectives.}
\label{fig:adjectives}
\vspace*{-0.5\baselineskip}
\vspace*{-2pt}
\end{figure}

To describe different teachers' teaching styles, we first observe the feedback questionnaires from parents to evaluate the teachers. The questionnaires are from TAL Education Group which contains over 10,000 surveys. And then, the adjectives about teaching styles are selected based on a Chinese text processing tool Thulac \cite{sun2016thulac}. As a result, 505 words are obtained, and through the manually selection, we remove those adjectives that not often used to describe teaching styles and finally gain the word list containing 41 teaching style adjectives. Furthermore, to have a more concise and clear description of teaching style, we separate 41 teaching style adjectives into four major teaching style categories, where the teaching styles in each major category have similar verbal and nonverbal behaviours. The details of the 41 teaching style adjectives and four major categories are shown in Figure \ref{fig:adjectives}.

\begin{figure}[t]
  \centering
  {
    \includegraphics[width= 0.40\textwidth]{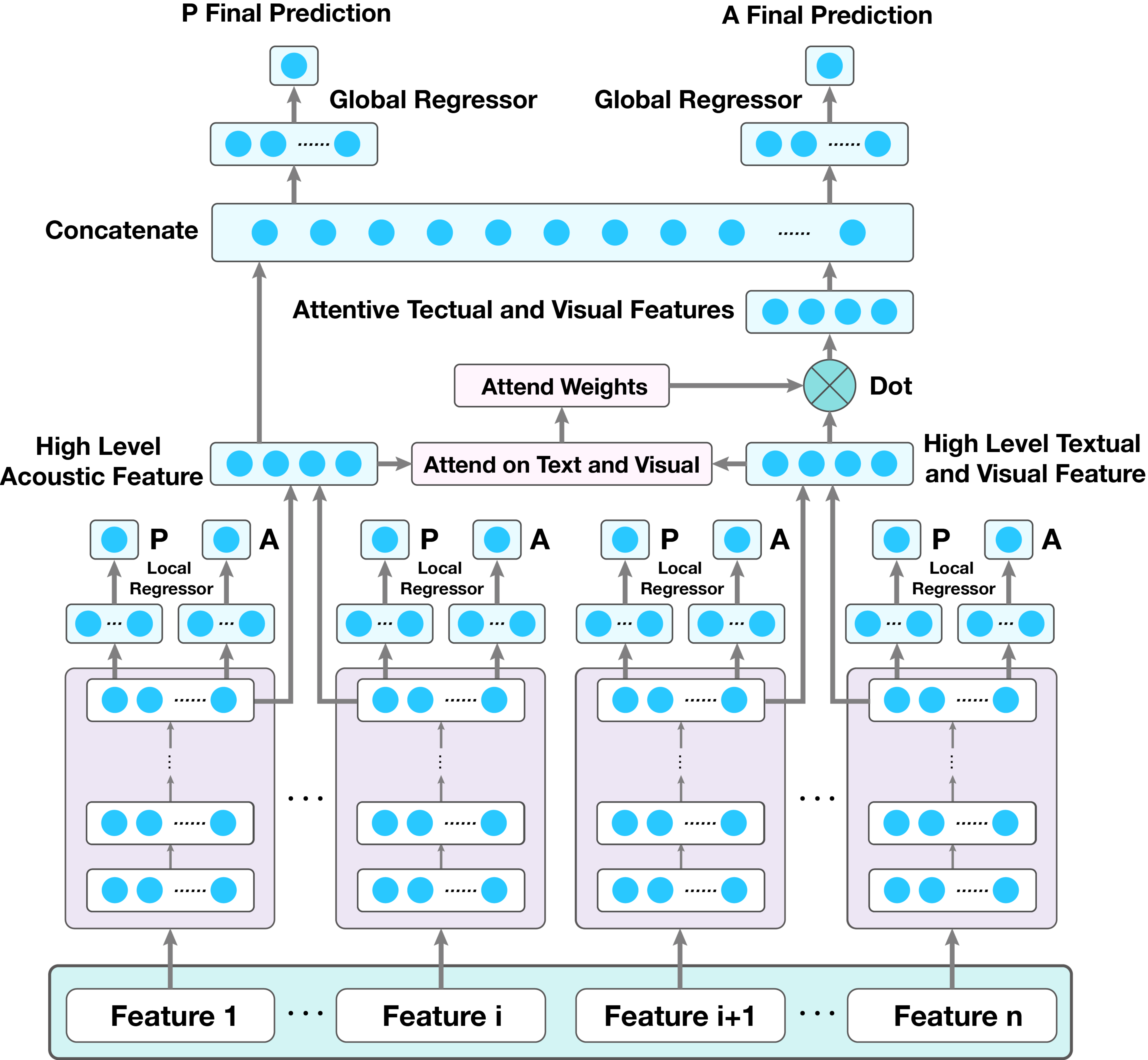}
  }
  \vspace*{-0.5\baselineskip}
  \caption{The structure of AMMDNN}
  \label{fig:AMMDNN}
  \vspace*{-1\baselineskip}
\vspace*{-2pt}
\end{figure}

Then we need to assign two-dimensional coordinates for each adjective. 
First, five different annotators are also invited to label the utterances with multiple teaching style adjectives. When more than three annotators choose an adjective for one utterance, this utterance is mapped to this adjective. Finally, for one teaching style adjective, it has a set of utterances mapping to it. The labelling methods of pleasure and arousal values for each utterance are presented in Section 5.2 and Table \ref{tab:4_pad_questions}. As we see, the center point of all the utterances is mapped to the teaching style adjective as its pleasure and arousal value. 

With this way, we can build the Teaching Style Semantic Space shown in Figure \ref{fig:tsss}. Specifically, the horizontal axis refers to pleasure value while the vertical axis refers to arousal value. Each point in the figure represents a teaching style adjective. We can see those positive adjectives such as humor, lively and friendly lay in the top right while negative adjectives like impatient, rigid and stiff lay in the bottom left.

\subsection{Attention-based Multi-path Multi-task Deep Neural Network}

\textbf{Intuition.} For traditional cross-media dimensional modelling, the entire features are employed as input and trained in a single regressor, which causes a high feature dimension. Thus, for a limited labeled training data, it would restrict the prediction performance to a great extent. Meanwhile, multi-task learning can improve generalization of a model by learning from related tasks \cite{chen2017multimodal}\cite{zhang2016cross}. Therefore, two strategies are applied to resolve our issue: 1) We adopt an attention-based Multi-path Deep Neural Network (MDNN) \cite{zhou2018inferring} to capture the internal correlations between cross-media features.First, it trains raw features from groups in local regressors to avoid high dimensions. Then high-level features of each local regressors of different modalities are concatenated based on attention mechanism as the input of a global regressor. More importantly, both local and global regressors are trained at the same time through a single objective function. 2) Intuitively pleasure and arousal which we predicted have close relations. Therefore, we promote the AMDNN to a novel structure named Attention-based Multi-task Multi-path Deep Neural Network (AMMDNN), shown in Figure \ref{fig:AMMDNN}.

\textbf{The structure of AMMDNN.} Instead of learning a single regressor with the whole sample features, the raw features are divided into small groups to learn multiple regressors based on different low-level descriptors (LLDs) and statistical functions, such as mean or standard deviation of MFCC features \cite{zhou2018inferring}. Therefore, each feature of different modalities can be used to train the responding regressor, which is called \emph{local regressor}. With the approach, the problem of high-dimensional inputs can be effectively avoided.  

It is noteworthy that although \emph{local regressors} take the independent nature of features into account, they largely ignore the relationships between different groups and modalities. To solve this issue, we merge the highest hidden layers of each \emph{local regressors} to generate a global representation based on the attention mechanism. Specifically, to better take advantages of the visual information and textual information, we adopt the attention mechanism mentioned in \cite{chen2018twitter} and modify the highest hidden layers to concatenate high-level representation features.

\begin{table}[t]
\centering
\begin{tabular}{c|c}
\hline
Notation&Definition\\
\hline\hline
${a}$ & the high-level acoustic feature\\
\hline
${w}$ & \tabincell{c}{the concatenated high-level visual and\\ textual feature}\\
\hline
${w_i}$ & the i-th dimensional feature of ${w}$， \\
\hline
${m}$ & the dimension number of ${w}$ \\
\hline
 ${f_a( \cdot ,a )}$ & attention function\\
\hline
${v}$ & the attentive visual and textual feature\\
\hline
${v_i}$ & the i-th dimensional feature of ${v}$\\
\hline
${x}$ &  the concatenated vector of ${a}$ and ${v}$ \\
\hline
$y$  &  the ground-truth label \\
\hline
$\hat{y}$  &  the prediction result \\
\hline
$T$  &  the number of tasks \\
\hline
$N$  &  the number of \emph{local regressor} \\
\hline
$\mathcal{L}_{t\textrm{,}g}$  &  \tabincell{c}{the cost function for \emph{global regressor} \\and the t-th task} \\
\hline
$\mathcal{L}_{t\textrm{,}l\textrm{,}n}$  &  \tabincell{c}{the cost function for n-th \emph{local regressor}\\  and t-th task}   \\
\hline
$\lambda$  &  the weight coefficient that between 0 and 1 \\
\hline
\end{tabular}
\caption{Notations and Definitions.}
\label{tab:notation}
\vspace*{-2\baselineskip}
\vspace*{-2pt}
\end{table}

For clarity, Table \ref{tab:notation} illustrates several
important notations and their definitions used in the paper.
${w_i}$, ${i \in(1,m)}$ represents the i-th dimensional feature of concatenated high-level visual and textual feature ${w}$, and ${f_a( \cdot ,a )}$ denote the attention function conditioned on the current high-level acoustic feature${a}$. ${v_i}$ represent the i-th dimensional feature of the attentive visual and textual feature ${v}$. Then, the attention weight ${\alpha_i}$ and attentive visual and textual feature ${v_i}$ is formulated as follows:
\vspace*{-0.5\baselineskip}
\begin{equation}
    u_i = f_a(w_i,  a)
\label{eq:1}
\vspace*{-0.5\baselineskip}
\end{equation}

\begin{equation}
    \alpha_i =\frac{exp(u_i)}{\sum_{i=1}^{m}exp(u_i)}
\label{eq:2}
\end{equation}
\begin{equation}
   v_i = \alpha_i \cdot w_i
\label{eq:3}
\end{equation}

A fully-connected layer with Exponential Linear Units (ELU) \cite{clevert2015fast} activation is chosen as the attention function and 
the attention vector ${v}$ is concatenated with
the high-level acoustic feature ${a}$ as the new input of the global regressor. 
Thus the concatenated high-level representation vector ${x}$ becomes ${[a, v]}$. 
Then, this representation is applied to train a \emph{global regressor}, which can effectively detect the correlations between different groups of features. Moreover, we optimize the framework through the single objective function, as shown in Equation \ref{eq:5}. The \emph{local regressors} and \emph{global regressor} are trained simultaneously with the function.

Meanwhile, \cite{chen2017multimodal} and \cite{zhang2016cross} show that multi-task learning can improve  model generalization by learning from related tasks. Intuitively predicting pleasure and arousal values have close relations\cite{chen2017multimodal}. Therefore, we can utilize multi-task learning to predict pleasure and arousal values together. Specifically, each regressor (both local and global ones) has two prediction tasks.

The mean square error (MSE) is applied as the loss function, which minimizes:
\vspace*{-0.5\baselineskip}
\begin{equation}
\label{eq:4}
\begin{aligned}
    \mathcal{L} = (\hat{y}-y)^{2}
\end{aligned}
\end{equation}
where $y$ is the ground-truth label,  $\hat{y}$ is the prediction result.
Therefore, our total loss function for the Multi-path Multi-task Deep Neural Network is as follows:
\vspace*{-0.5\baselineskip}
\begin{equation}
\label{eq:5}
\begin{aligned}
    \mathcal{L}
    &= \sum_{t=1}^{T}((1-\lambda)\mathcal{L}_{t\textrm{,}g} +  \lambda\sum_{n=1}^{N}{ \mathcal{L}_{t\textrm{,}l\textrm{,}n}})\\
\end{aligned}
\end{equation}

Across all tasks $T$, $\mathcal{L}_{t\textrm{,}g}$ is the cost function for \emph{global regressor} and the t-th task while $\mathcal{L}_{t\textrm{,}l\textrm{,}n}$ is the cost function for n-th \emph{local regressor} and t-th task. $\lambda$ is the weight coefficient that between 0 and 1(we set $\lambda$=0.5 in our experiment). $N$ is the number of \emph{local regressor} and $T$ is the number of our tasks.

As shown in Figure \ref{fig:AMMDNN}, the motivation of this acoustic-guide Attention-based Multi-path Multi-task Deep Neural Network is that we use the acoustic features to guide the attention weights of the textual features and visual features in order to enforce the model to self-select which dimension feature it should attend on.

\subsection{Mapping Coordinates with Teaching Styles.}

Based on the proposed AMMDNN, we can get two-dimensional coordinates for each utterance. Then we select the adjective word on the TSSS with the shortest Euclidean distances for the coordinates. In this way, we can obtain three teaching style words that fit with the utterance most.

\section{Experiments}

\begin{table}[t]
\centering
\begin{tabular}{c|c|c|c|c}
\hline
  Regression & P RMSE & P CCC & A RMSE & A CCC\\
\hline
DNN & 0.844 & 0.554 & 0.722 & 0.669\\
\cline{1-5}
 SVM & \textbf{0.680} & 0.703 & \textbf{0.565} & 0.787 \\
\cline{1-5}
 DTree & 1.006 & 0.511 & 0.834 & 0.623\\
\cline{1-5}
RF & 0.715 & 0.672 & 0.596 & 0.760\\
\cline{1-5}
 MDNN & 0.707 & 0.694 & 0.648 & 0.765\\
\cline{1-5}
 AMDNN & 0.710 & 0.729  & 0.587 & 0.805\\
\cline{1-5}
AMMDNN  & 0.713 & \textbf{0.739} & 0.582 & \textbf{0.808}\\
\cline{1-5}
\end{tabular}
\caption{Comparison among different regression models for arousal and pleasure prediction.}
\label{tab:performance}
\vspace*{-2\baselineskip}
\vspace*{-2pt}
\end{table}

\begin{figure}[t]
\centering
\includegraphics[width=0.45\textwidth]{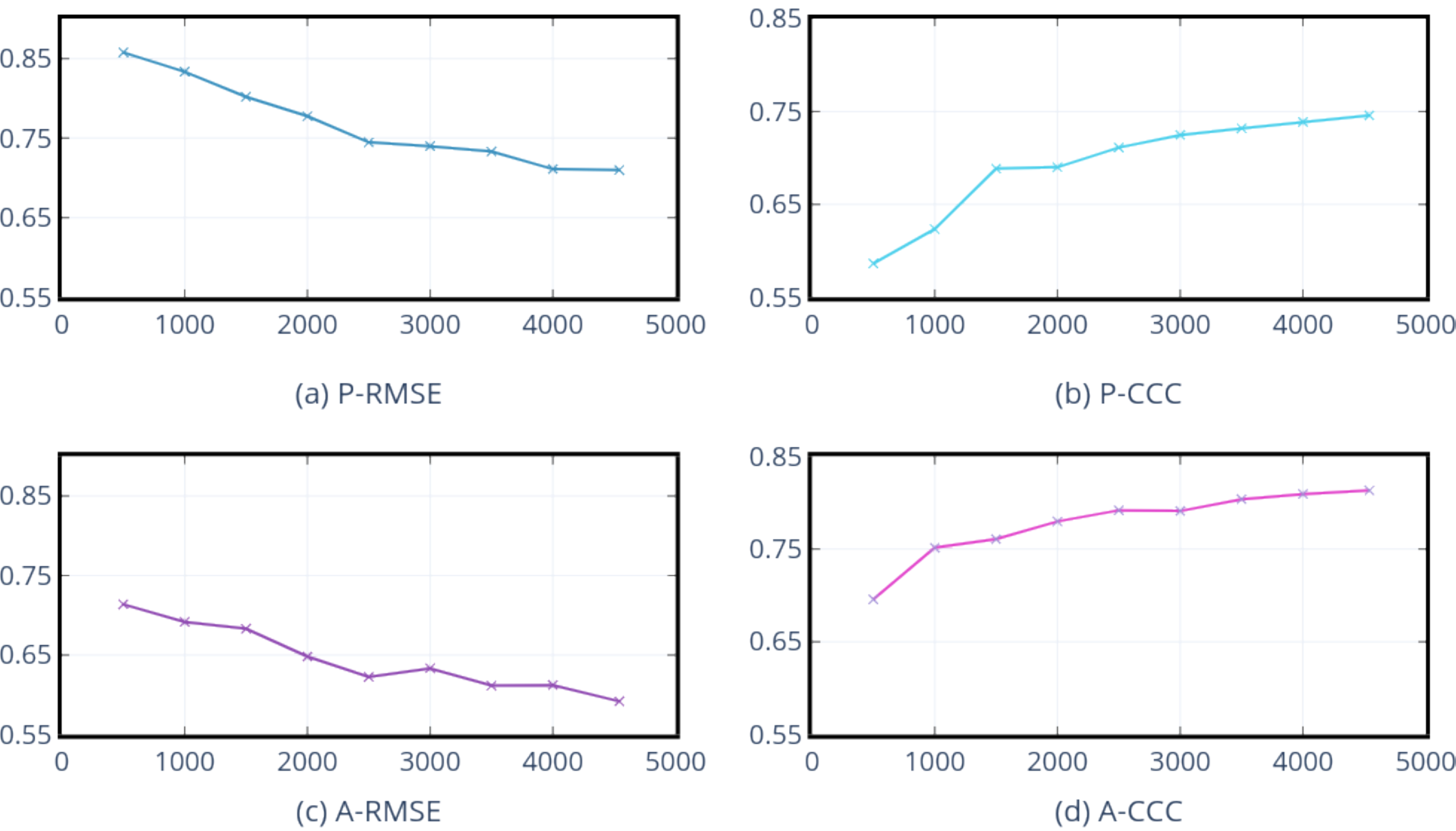}
\vspace*{-0.5\baselineskip}
\caption{Training data scalability analysis.}
\label{fig:train_size}
\vspace*{-0.5\baselineskip}
\vspace*{-2pt}
\end{figure}

\subsection{Data Collection}
\begin{figure}[t]
\centering
\includegraphics[width=0.45\textwidth]{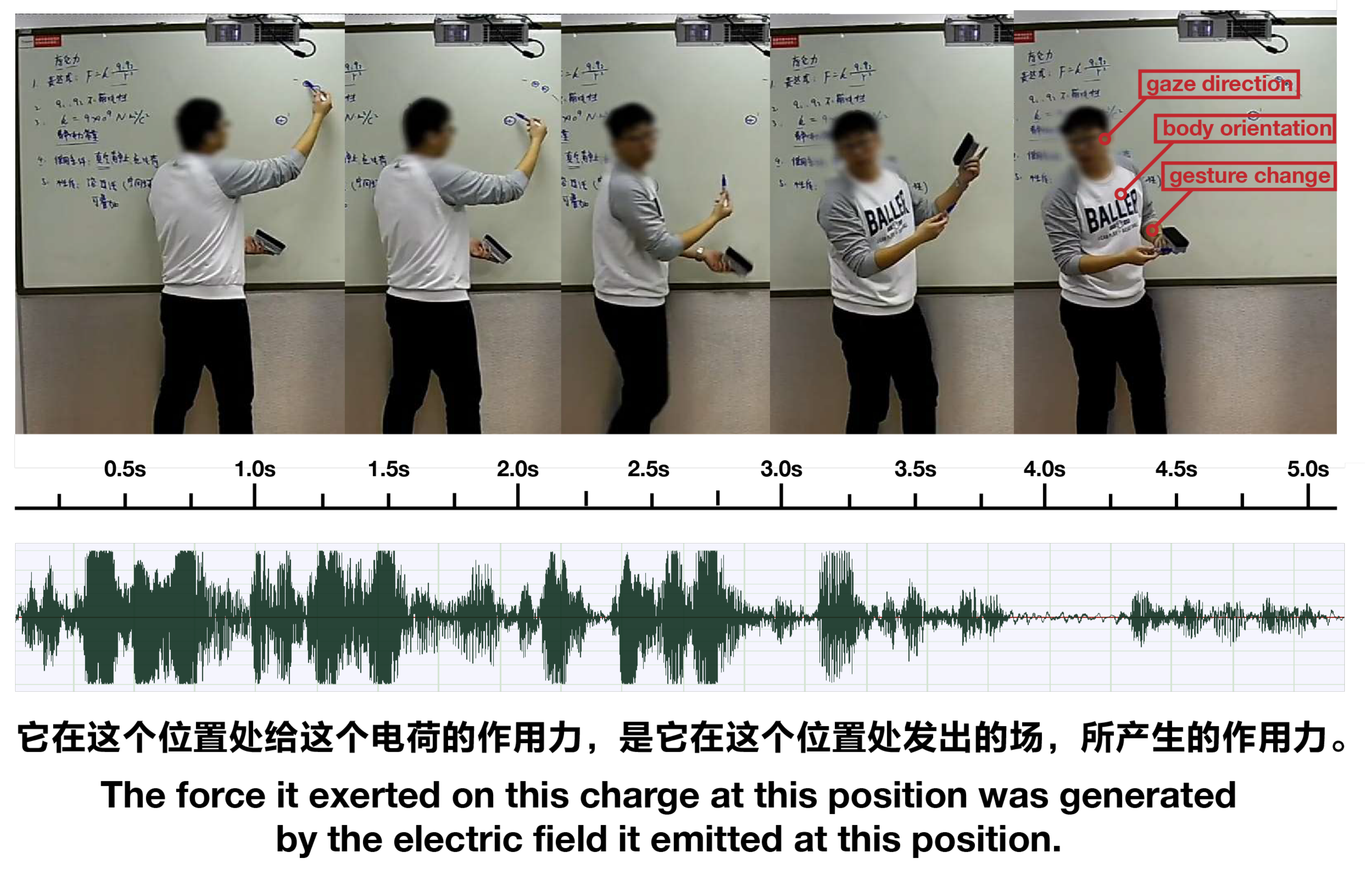}
\vspace*{-1\baselineskip}
\caption{An example of data in TEG18.}

\label{fig:sample_data}
\vspace*{-0.5\baselineskip}
\vspace*{-2pt}
\end{figure}

\textbf{TEG18.}
We build a sizeable full-annotated benchmark dataset, which entirely contains 4,541 cross-media utterances recorded in the educational environment from TAL Education Group(TEG18). All of the cross-media utterances are recorded from the real primary school classes. Data samples are randomly collected from different teachers and different courses. The dataset covers both male and female teachers. Specifically, there are 42 teachers in total, and 24 are female while 18 are male. Also, each of the cross-media utterances is 10 seconds long. An example of data in TEG18 is shown in Figure \ref{fig:sample_data}.
, and the main contents are teachers' speeches.

\subsection{Labeling}
In terms of pleasure and arousal labeling, based on the PAD model proposed by \cite{mehrabian1996pleasure}, \cite{li2005reliability} provides a Chinese version questionnaire to evaluate pleasure, arousal, and dominance values. Specifically, annotators need to answer four different questions for each value. Then, the paper gives three formulas for calculating the three values.

For our work, considering the data are recorded in the teaching environment, the dominance of teachers will be high, we annotate the teaching styles adjectives with pleasure and arousal labels. Based on the Chinese version questionnaire, we make some changes and reduce the number of questions to eliminate the ambiguity and make the survey more suitable for the education environment. As is shown in Table \ref{tab:4_pad_questions}, we design four questions for annotators to answer. Specifically, question 1 and 4 are about arousal value and question 2 and 3 are about pleasure value. For each question, there are two different descriptions indicating -2 and 2 respectively. Annotators need to choose a number between -2 and 2. Then, we normalize all the labeled values to make the mean 0 and the variance 1.
\vspace*{-0.5\baselineskip}
\begin{equation}
\label{eq:6}
    P=-Q_2+Q_3
    \vspace*{-0.5\baselineskip}
\end{equation}
\vspace*{-0.5\baselineskip}
\begin{equation}
\label{eq:7}
    A=-Q_1+Q_4
    \vspace*{-0.5\baselineskip}
\end{equation}

\begin{table}[t]
\centering
\begin{tabular}{c|c|c|c|c|c|c}
\hline
&-2&-1&0&1&2&\\
\hline
Wide awake &&&&&& Sleepy\\
\hline
Friendly &&&&&& Strict\\
\hline
Harsh Sound &&&&&& Comfortable Sound\\
\hline
Cautious &&&&&& Surprising\\
\hline
\end{tabular}
\caption{Four questions for the labelers to answer.}
\label{tab:4_pad_questions}
\vspace*{-2\baselineskip}
\vspace*{-2pt}
\end{table}
 
Each label of every utterance is annotated by four different annotators independently. To access inter-rater reliability, we calculate the Cronbach alpha coefficients for all labelers regarding pleasure and arousal values. 
The Cronbach alpha coefficients are 0.782 for pleasure and 0.828 in terms of arousal values.
Compared with the public PAD full-annotated dataset IEMOCAP \cite{busso2008iemocap}, where Cronbach alpha coefficients are 0.809 and 0.607 in terms of pleasure and arousal values, the results prove the effectiveness of our dataset.

\subsection{Feature Extraction}

\textbf{Acoustic feature.}
 We utilize openSMILE toolkit \cite{eyben2010opensmile} to extract acoustic features for TEG18. Totally, we obtain 1,582 statistic acoustic features, which is the same as the acoustic features used in the INTERSPEECH 2010 Paralinguistic Challenge \cite{schuller2010interspeech}. 
 
\begin{figure}[t]
\centering
\includegraphics[width=0.40\textwidth]{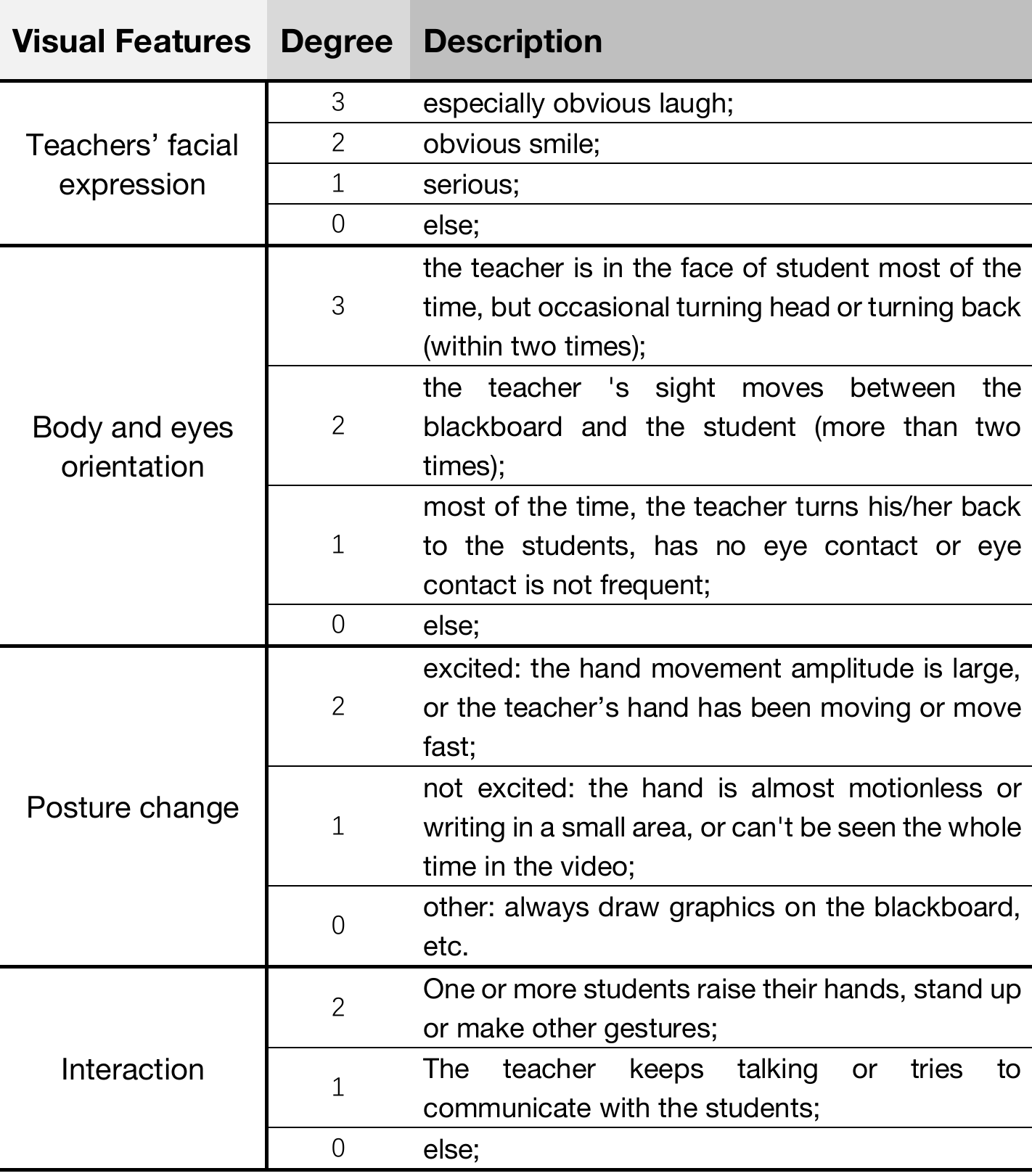}
\vspace*{-0.5\baselineskip}
\caption{Visual feature details.}
\vspace*{-1\baselineskip}
\label{fig:visual——feature}

\vspace*{-2pt}
\end{figure}
 
\textbf{Visual feature.} 
We define the visual features of teachers' teaching classes from four
aspects: degree of teachers' facial expression, degree of body and eyes orientation, degree of posture change, degree of interaction. We invite three people to annotate them, and the annotation details are listed in Figure \ref{fig:visual——feature}. Totally, we have 14-dimensional visual features.

\textbf{Textual feature.}
For textual information in TEG18, we first use Thulac Tool\cite{li2009punctuation} which is an efficient Chinese word segmentation to get words of an utterance. Then we utilize word2vec to learn word embeddings. Specifically, we use the whole 31.2 million chinese word corpora collected from the 7.5 million utterance from SVAD13\cite{zhou2018inferring} as the training corpora for word2vec. 
Then, we extract 4200-dimensional utterance-level textual features according to the statistic functions (mean, std, disp,  max,  min,  range,  quartile1/2/3,  iqr1-2/2-3/1-3, skewness, kurtosis) over the LLDs.

\subsection{Experimental Setup}
\textbf{Evaluation metrics.} In all the experiments, we evaluate the performance in terms of concordance correlation coefficient(CCC). The results reported in TEG18 are based on five-fold cross validation.
\textbf{Comparison methods.}
For predicting arousal and pleasure values, we utilize six different baseline regression models: Random Forest (RF), Support Vector Machine (SVM), Decision Tree (DTree) \cite{Trendowicz2014Classification},
Deep Neural Network (DNN) \cite{Bengio2009Learning}, Multi-path Deep Neural Network (MDNN)\cite{zhou2018inferring}, and Attention-based Multi-path Deep Neural Network(AMDNN).

\textbf{Construction setting.}
In implementation of comparisons, we set $\lambda$=0.5, $N$ = 7 in Eq. \ref{eq:5} for the comparison methods utilize multi-path structure. Each local regressor contains two hidden layers with 400 units. Dropout \cite{srivastava2014dropout} is applied for each hidden layer with a dropout ratio of $0.5$. The optimization method we adopt is Adam \cite{kingma2014adam} with an initial learning rate at $10^{-4}$. And the activation function we apply is ELU activation.

\subsection{Performance}

To evaluate the effectiveness of our proposed Multi-path Multi-task Deep Neural Network(AMMDNN), we compare the performance with some baseline methods with cross-media information.
Table \ref{tab:performance} shows the results.
In terms of CCC, the proposed AMMDNN outperforms all the baseline methods: for arousal, +0.185 compared with DTree, +0.048 compared with RF and +0.021 compared with SVM. For pleasure, +0.228 compared with DTree, +0.067 compared with RF and +0.036 compared with SVM.
Specifically, 1) the MDNN outperforms the DNN (+0.096) in terms of CCC with arousal and (+0.140) in terms of CCC with pleasure. This proves the effectiveness of the multi-path component which considers the in-dependency nature of different features and avoid high-dimensional inputs with limit training data . 
2) the AMDNN outperforms the MDNN (+0.040) in terms of CCC with arousal and (+0.035) in terms of CCC with pleasure. This proves the effectiveness of the proposed attention mechanism, which utilize the acoustic features to guide the attention weights of the textual features and visual features to help enforce the model to self-select which dimension feature it should attend on. 
3) the AMMDNN outperforms the AMDNN (+0.003) in terms of CCC with arousal and (+0.010) in terms of CCC with pleasure. This proves the effectiveness of the proposed multi-task method which consider predicting pleasure and arousal values as related tasks.

\subsection{Analysis}
\begin{figure}[t]
\centering
\includegraphics[height=38mm,width=0.9\columnwidth]{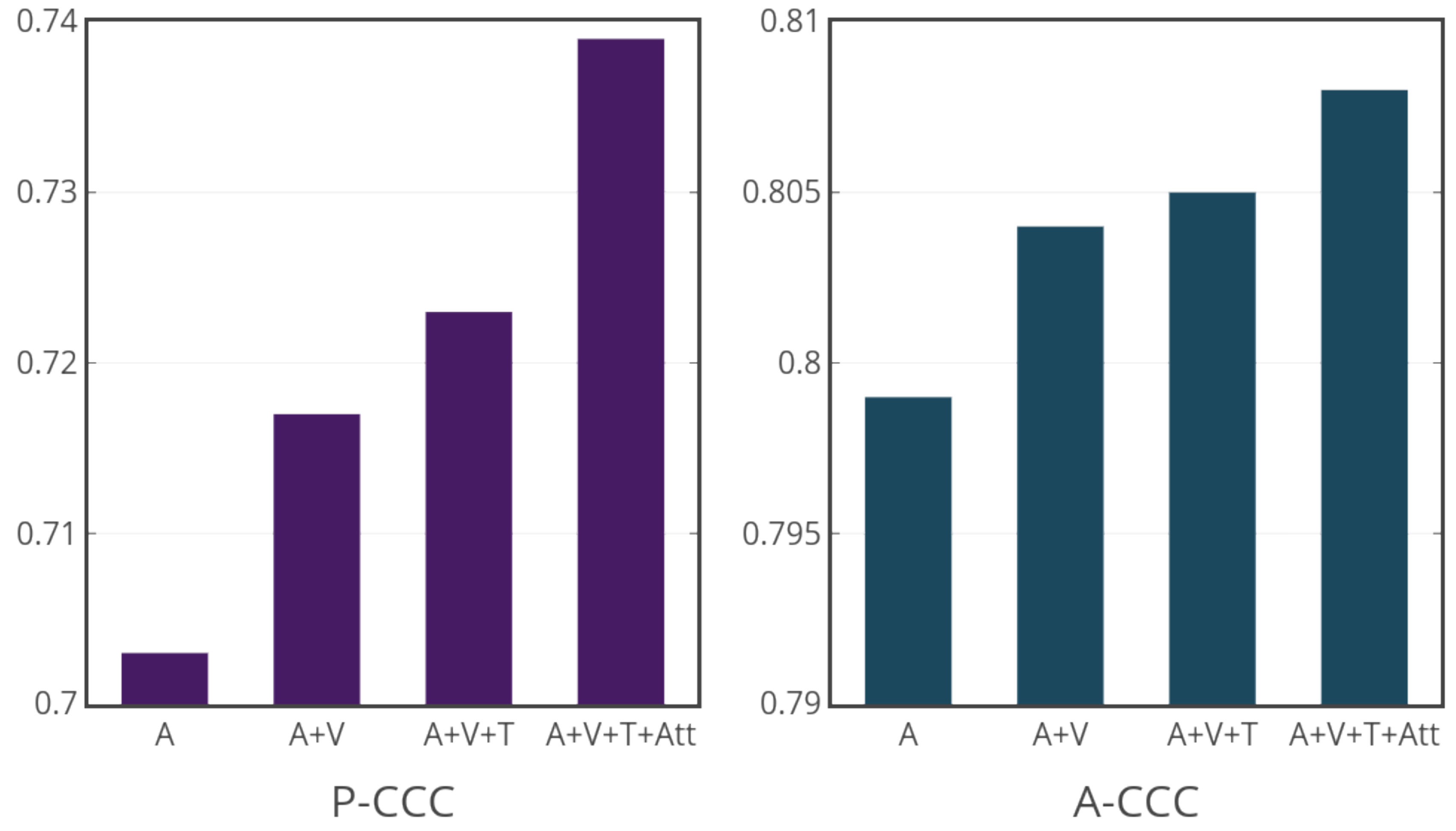}
\caption{Feature contribution analysis.\\
(A:Acoustic, V:Visual, T:Textual, Att:Attention)}
\vspace*{-0.5\baselineskip}
\label{fig:feature_contrbution}
\vspace*{-0.5\baselineskip}
\vspace*{-2pt}
\end{figure}

\textbf{Feature contribution analysis.}
We first discuss the contributions of acoustic, visual and textual features in understanding teaching styles. Specifically, for `Acoustic Only', `Acoustic+Visual', `Acoustic+Visual+ Textual', we utilize Multi-path multi-task Deep Neural Network model. And for `Acoustic+Visual+Textual', we utilize AMMDNN model. The CCC results for arousal and pleasure are shown in Figure \ref{fig:feature_contrbution}. As show in the figure, the "Acoustic+Visual" outperforms the "Acoustic Only" (+0.014) in terms of CCC with pleasure and (+0.005) in terms of CCC with arousal. This validates the necessity of taking the textual information into consideration. The "Acoustic+Visual+ Textual" outperforms the "Acoustic+Visual" (+0.006) in terms of CCC with pleasure and (+0.001) in terms of CCC with arousal, which validates the necessity of taking the textual information into consideration. The "Acoustic+Visual+Textual+Attention" outperforms the "Acoustic+Visual+\\Textual" (+0.016) in terms of CCC with pleasure and (+0.003) in terms of CCC with arousal, which indicates that our proposed attention mechanism can be more effective in modelling multi-modal features. 

\textbf{Parameter sensitivity analysis.}
We further test the parameter sensitivity about training data size. From Figure \ref{fig:train_size}, we can find that as the scale of training data increases, performance obviously gets better for all of the evaluation metrics (pleasure RMSE, pleasure CCC, arousal RMSE, and arousal CCC). Specifically, the pleasure has a higher improvement than arousal. The growth trend is slowing down after training size reaches 2500. With the size over 4500, the performance almost reaches convergence. Considering the difficultly for manually labelling, we choose 4541 as our dataset.

\begin{figure*}[t]
\centering
\includegraphics[width=1.0\textwidth]{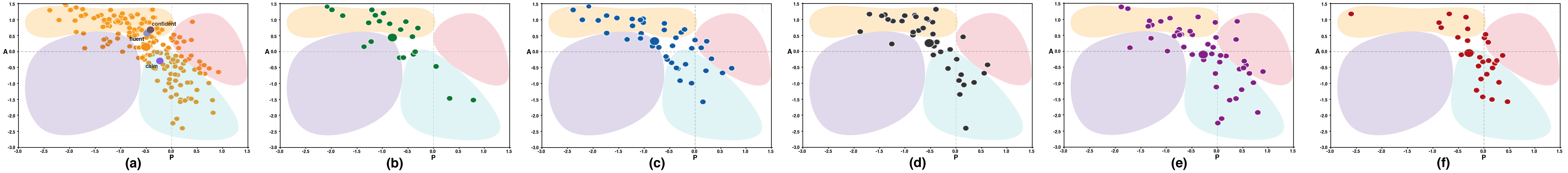}
\vspace*{-0.5\baselineskip}
\caption{Physics course styles analysis.}
\label{fig:dif_course1}
\vspace*{-0.5\baselineskip}
\vspace*{-2pt}
\end{figure*}
\begin{figure*}[t]
\centering
\includegraphics[width=1.0\textwidth]{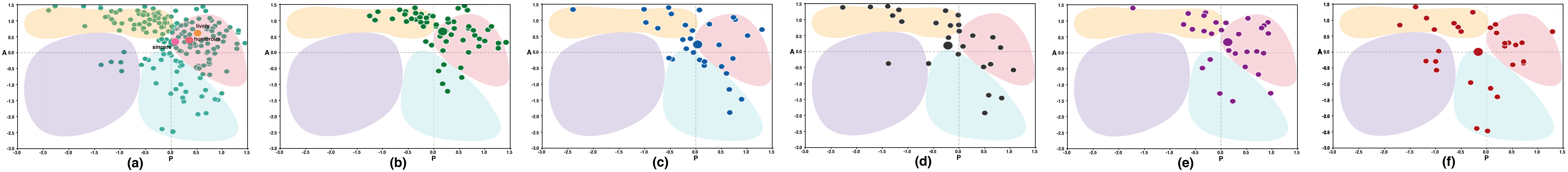}
\vspace*{-0.5\baselineskip}
\caption{Chinese course styles analysis.}
\label{fig:dif_course2}
\vspace*{-0.5\baselineskip}
\vspace*{-2pt}
\end{figure*}
\begin{figure}[t]
\centering
\includegraphics[width=0.45\textwidth]{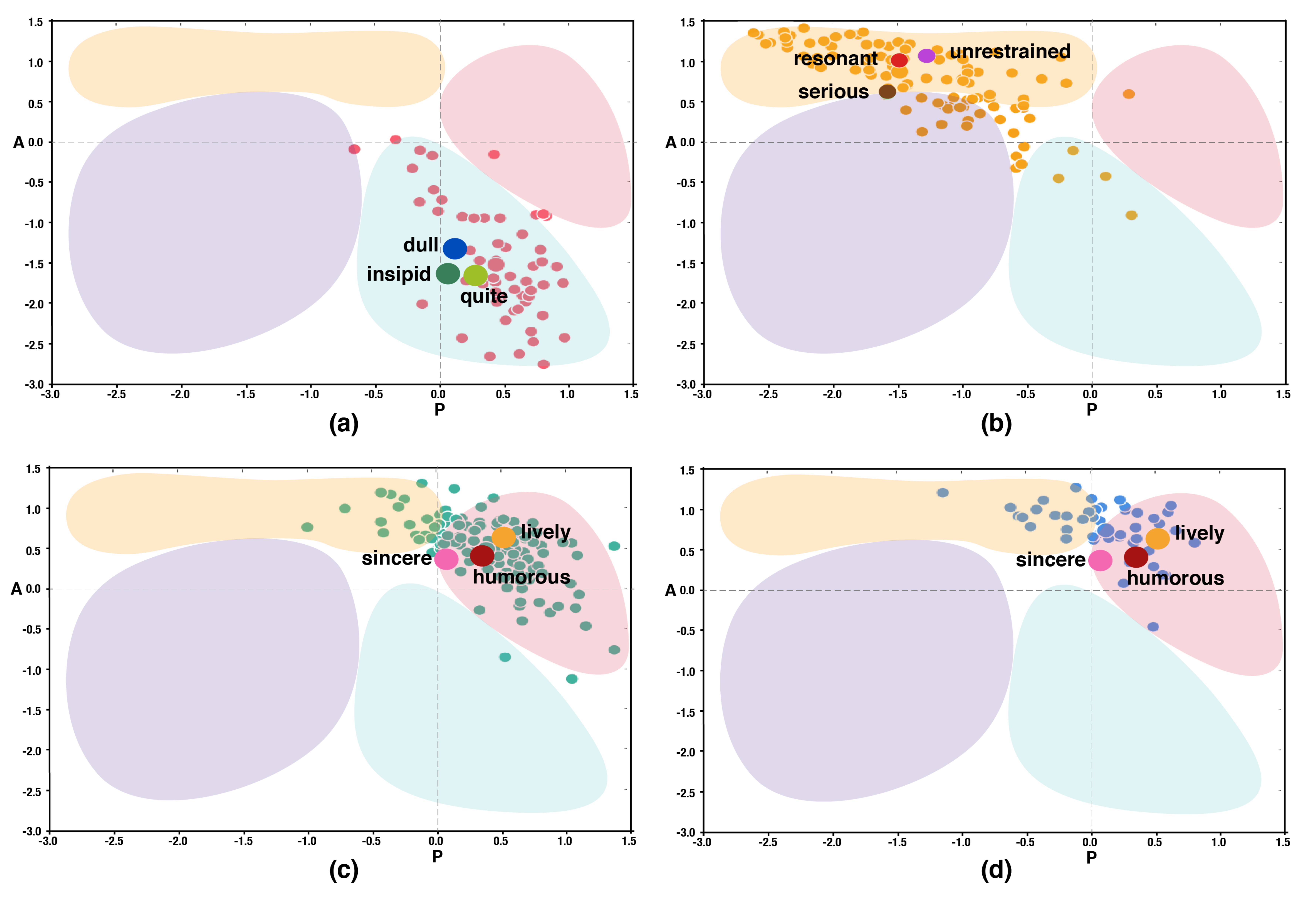}
\vspace*{-0.5\baselineskip}
\caption{Personal teaching styles analysis.}
\label{fig:dif-teacher_all}
\vspace*{-0.5\baselineskip}
\vspace*{-2pt}
\end{figure}

\subsection{Case Study}

With the proposed AMMDNN, we are capable of mapping cross-media features to two-dimensional coordinates on the TSSS. Then, we select the adjective word on the TSSS with the shortest Euclidean distances for the coordinates so that we can get the most suitable teaching style word for each utterance. Based on our prediction results for teaching style adjectives, we conduct some interesting case studies to further show the advantages and universality of our solution.

\textbf{Teaching styles comparison among different teachers.} We can establish a personal TSSS for every teacher to analyze his/her teaching styles. To determine the teaching styles of each teacher, first, we apply AMMDNN to calculate the pleasure and arousal values of an utterance set corresponding to the teacher. Then, we assign an coordinate value for the teacher by calculating the center of gravity of the points in the utterance set. Finally, we choose three teaching style words which have the shortest Euclidean distance with our calculated teacher coordinate value in TSSS. Therefore, the three chosen teaching style words are considered as the teaching style for the teacher.

We randomly choose two male teachers shown in Figure \ref{fig:dif-teacher_all}.(a) and (b) and two female teachers shown in Figure \ref{fig:dif-teacher_all}.(c) and (d).
Based on the TSSS, the teacher in Figure \ref{fig:dif-teacher_all}.(a) may have the teaching styles of insipid, dull and quite,  since he has high pleasure values and quite low arousal values.
In terms of the teacher in Figure \ref{fig:dif-teacher_all}.(b), this teacher has low pleasure and high arousal values. Therefore, he may have the teaching styles of serious, resonant and unrestrained.
Regarding the teacher in Figure \ref{fig:dif-teacher_all}.(c), she has a wide range of the pleasure and arousal values, while the values are all not very high, indicating that she may have sincere, humorous and lively teaching styles.
The teacher in Figure \ref{fig:dif-teacher_all}.(d) may be sincere, humorous and lively, since she has high pleasure and arousal values. Based on the TSSS, we can better understand teachers' teaching styles and analyze them objectively and give them quantitative feedbacks.

\textbf{Teaching styles comparison among different courses.} We further analyze teaching styles among different courses. We apply the proposed AMMDNN to predict both pleasure and arousal values for different courses. By drawing the pleasure-arousal values on our TSSS for each course, a variety of teaching style features among different courses is revealed evidently. Take physics as an example, it (Figure \ref{fig:dif_course1}, graph(a)) appears a considerable tendency toward lower pleasure value when compared with Chinese(Figure \ref{fig:dif_course2}, graph(a)), which indicates that in a class for science subjects such as physics, the teacher generally acts more seriously than in a class for subjects that do not put forward such a strong request for scientific thinking. 

Furthermore, we also analysis the change of teaching style over different periods in a class for certain subject. We divide each class into five segments by time equally. Then we draw the data in different segment separately in our TSSS, as shown in Figure \ref{fig:dif_course1}.(b-f) and \ref{fig:dif_course2}.(b-f). For each segment, we also compute the average pleasure and arousal value and show it on our TSSS. 
And the outcome graphs display distinct teaching style changes by time in different courses. Again, the physics course is treated as an example. As shown in \ref{fig:dif_course1}.(b-f), with the timeline of a physics class proceeds, the pleasure value start to arise and arousal value start to shrink. And that phenomenon sits well with the common sense that in such a scientific class which is usually accused of excessive seriousness, the teacher gradually becomes more exhausted and less enthusiastic than at the beginning of the class.

\begin{figure}[t]
\centering
\includegraphics[width=0.45\textwidth]{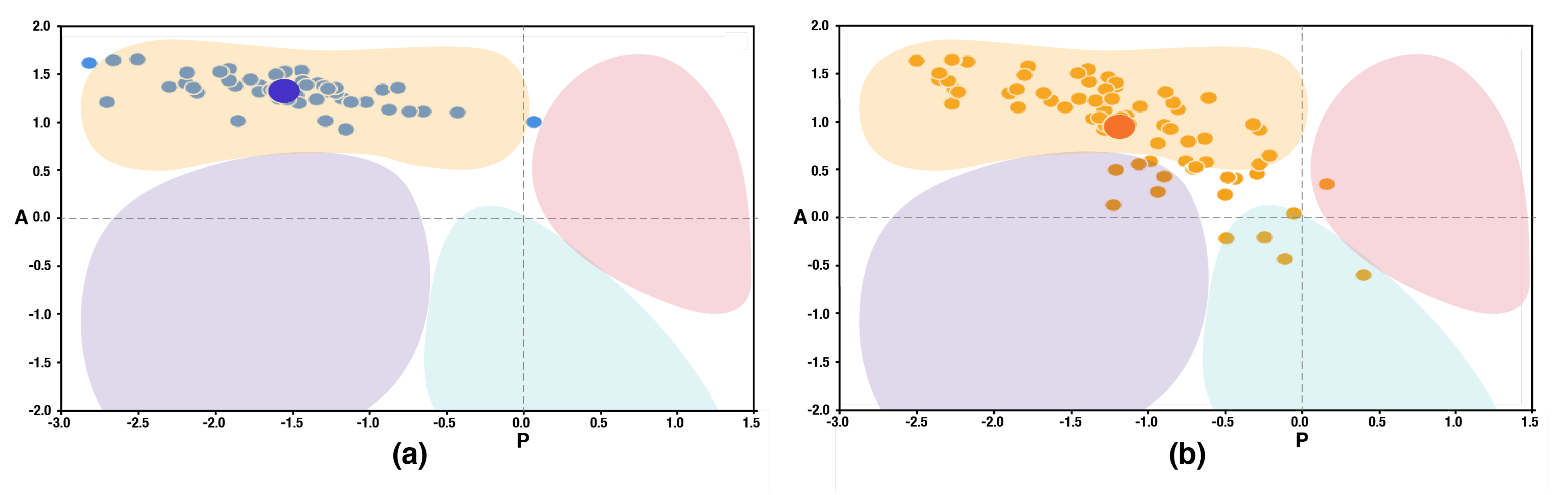}
\caption{Teaching styles analysis among different attention rate for lesson(I).}
\label{fig:dif_lesson1}
\vspace*{-0.5\baselineskip}
\vspace*{-2pt}
\end{figure}
\begin{figure}[t]
\centering
\includegraphics[width=0.45\textwidth]{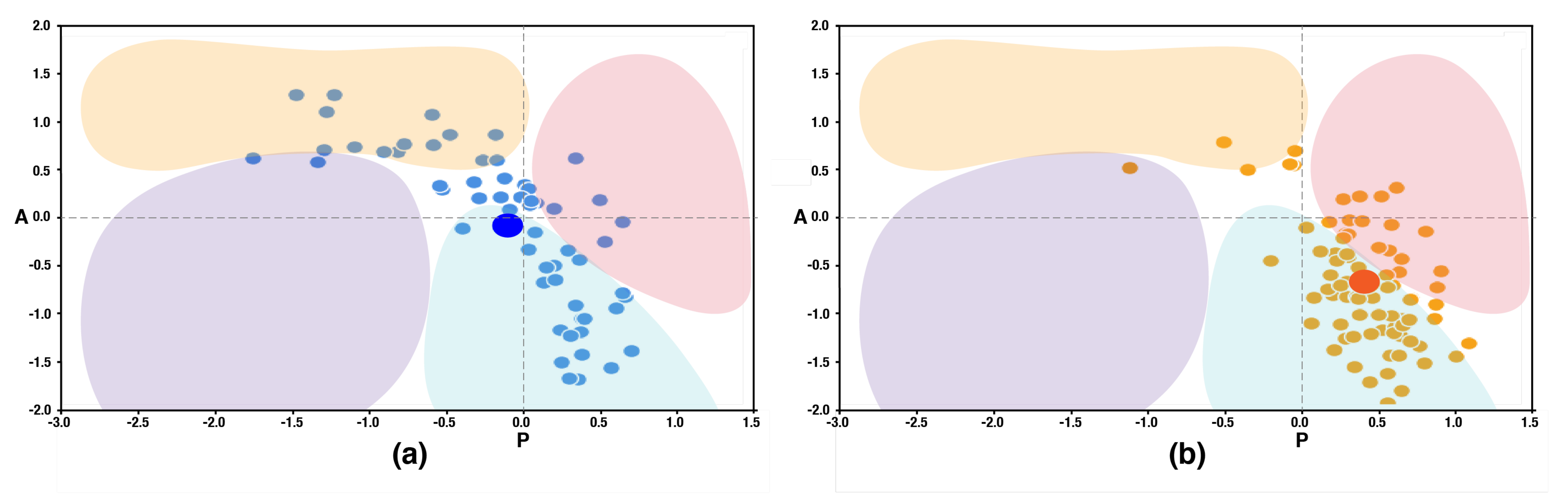}
\caption{Teaching styles analysis among different attention rate for lesson(II).}
\label{fig:dif_lesson2}
\vspace*{-0.5\baselineskip}
\vspace*{-2pt}
\end{figure}

\textbf{Teaching styles analysis among different attention rate during a class.} We calculate students' concentration on teachers during a class and analyze the results with the predicted pleasure and arousal values. 

First, we calculate how much the students pay attention to the teacher. For each frame of the teaching class video, we calculate the students' head position coordinates and the students' line of sight vector in the same coordinate system. First, the head coordinates and the line of sight vector are projected onto a two-dimensional $x\-y$ plane. Second, we calculate all intersections of the students' line of sight and record the number of intersections as $n$. Third, we sort the $x$ coordinate values of the intersections and calculate the average after removing the largest $n/4$ values and the smallest $n/4$ values. This average is regarded as the predicted $x$ coordinate of the teacher. Similarly, we can get the predicted teacher's $y$ coordinate. Finally, we get the distance from the predicted position of the teacher on the $x\-y$ plane to the line of sight for every student. 
Then, we divide them into three categories(category I, II, III) according to the distance mention above. Specifically, scores under half of the average are put into category I, which means a lack of attention. Scores more than twice the average are put into category II, which means the teacher is highly focused on. Otherwise, it will be put into category III.

We collect a dataset containing 39 lessons from real high school teaching classes provided by EduBrain.ai. Then we randomly choose two lessons and cut them into 10s utterances. Next we predict the pleasure and arousal values of these utterances utilizing our proposed model. As shown in Figure \ref{fig:dif_lesson1} and \ref{fig:dif_lesson2}, Figure \ref{fig:dif_lesson1}.(a) and \ref{fig:dif_lesson2}.(a) are utterances of category II, which have high student attention. Figure \ref{fig:dif_lesson1}.(b) and \ref{fig:dif_lesson2}.(b) are utterances of category I, which show low attention of students. We suggest that students pay more attention when teachers have lower pleasure values and higher arousal values in class. 
Therefore, teaching styles like passionate and severe which have low pleasure and high arousal values may catch more attention in class.

\section{Conclusion}
In this paper, we make an important step towards understanding teaching styles.
We build a Teaching Styles Semantic Space (TSSS) to describe teaching styles. Then, we select the most often used 41 teaching style adjectives from more than 10,000 feedback questionnaires provided by Tomorrow Advancing Life (TAL) Education Group. Further, we propose an Attention-based Multi-path Multi-task Deep Neural Network (AMMDNN) to map cross-media features to the coordinates on the TSSS. Our result shows that AMMDNN turns out to be a practical solution. Based on the TSSS, we can obtain the teaching style adjective word that have the shortest Euclidean distances with the coordinates so that we can finally map teachers' utterances with teaching styles. Moreover, we conduct some interesting case studies which show the effectiveness of the dimensional space for understanding teaching styles.

\begin{acks}
This work is guided by Mr. Yan Huang from TAL Education Group. We would like to sincerely thank TAL Education Group and Mr. Yan Huang for their support and guidance to the work. We would also like to thank EduBrain.ai and Shijiazhuang No.1 High School Hebei for their support.

This work is supported by National Key Research and Development Plan (2016YFB1001200),the Innovation Method Fund of China (2016IM010200), the state key program of the National Natural Science Foundation of China (NSFC) (No.61831022) and National Natural, and Science Foundation of China(61521002).
\end{acks}
\newpage
\bibliographystyle{ACM-Reference-Format}
\balance 
\bibliography{sample-sigconf}


\begin{thebibliography}{28}


\ifx \showCODEN    \undefined \def \showCODEN     #1{\unskip}     \fi
\ifx \showDOI      \undefined \def \showDOI       #1{#1}\fi
\ifx \showISBNx    \undefined \def \showISBNx     #1{\unskip}     \fi
\ifx \showISBNxiii \undefined \def \showISBNxiii  #1{\unskip}     \fi
\ifx \showISSN     \undefined \def \showISSN      #1{\unskip}     \fi
\ifx \showLCCN     \undefined \def \showLCCN      #1{\unskip}     \fi
\ifx \shownote     \undefined \def \shownote      #1{#1}          \fi
\ifx \showarticletitle \undefined \def \showarticletitle #1{#1}   \fi
\ifx \showURL      \undefined \def \showURL       {\relax}        \fi
\providecommand\bibfield[2]{#2}
\providecommand\bibinfo[2]{#2}
\providecommand\natexlab[1]{#1}
\providecommand\showeprint[2][]{arXiv:#2}

\bibitem[\protect\citeauthoryear{Atkins and Brown}{Atkins and Brown}{2002}]%
        {atkins2002effective}
\bibfield{author}{\bibinfo{person}{Madeleine Atkins} {and}
  \bibinfo{person}{George Brown}.} \bibinfo{year}{2002}\natexlab{}.
\newblock \bibinfo{booktitle}{\emph{Effective teaching in higher education}}.
\newblock \bibinfo{publisher}{Routledge}.
\newblock


\bibitem[\protect\citeauthoryear{Bengio}{Bengio}{2009}]%
        {Bengio2009Learning}
\bibfield{author}{\bibinfo{person}{Yoshua Bengio}.}
  \bibinfo{year}{2009}\natexlab{}.
\newblock \showarticletitle{Learning Deep Architectures for AI}.
\newblock \bibinfo{journal}{\emph{Foundations \& Trends® in Machine Learning}}
  \bibinfo{volume}{2}, \bibinfo{number}{1} (\bibinfo{year}{2009}),
  \bibinfo{pages}{1--127}.
\newblock


\bibitem[\protect\citeauthoryear{Busso, Bulut, Lee, Kazemzadeh, Mower, Kim,
  Chang, Lee, and Narayanan}{Busso et~al\mbox{.}}{2008}]%
        {busso2008iemocap}
\bibfield{author}{\bibinfo{person}{Carlos Busso}, \bibinfo{person}{Murtaza
  Bulut}, \bibinfo{person}{Chi-Chun Lee}, \bibinfo{person}{Abe Kazemzadeh},
  \bibinfo{person}{Emily Mower}, \bibinfo{person}{Samuel Kim},
  \bibinfo{person}{Jeannette~N Chang}, \bibinfo{person}{Sungbok Lee}, {and}
  \bibinfo{person}{Shrikanth~S Narayanan}.} \bibinfo{year}{2008}\natexlab{}.
\newblock \showarticletitle{IEMOCAP: Interactive emotional dyadic motion
  capture database}.
\newblock \bibinfo{journal}{\emph{Language resources and evaluation}}
  \bibinfo{volume}{42}, \bibinfo{number}{4} (\bibinfo{year}{2008}),
  \bibinfo{pages}{335}.
\newblock


\bibitem[\protect\citeauthoryear{Chen, Jin, Zhao, and Wang}{Chen
  et~al\mbox{.}}{2017}]%
        {chen2017multimodal}
\bibfield{author}{\bibinfo{person}{Shizhe Chen}, \bibinfo{person}{Qin Jin},
  \bibinfo{person}{Jinming Zhao}, {and} \bibinfo{person}{Shuai Wang}.}
  \bibinfo{year}{2017}\natexlab{}.
\newblock \showarticletitle{Multimodal multi-task learning for dimensional and
  continuous emotion recognition}. In \bibinfo{booktitle}{\emph{Proceedings of
  the 7th Annual Workshop on Audio/Visual Emotion Challenge}}. ACM,
  \bibinfo{pages}{19--26}.
\newblock


\bibitem[\protect\citeauthoryear{Chen, Yuan, You, and Luo}{Chen
  et~al\mbox{.}}{2018}]%
        {chen2018twitter}
\bibfield{author}{\bibinfo{person}{Yuxiao Chen}, \bibinfo{person}{Jianbo Yuan},
  \bibinfo{person}{Quanzeng You}, {and} \bibinfo{person}{Jiebo Luo}.}
  \bibinfo{year}{2018}\natexlab{}.
\newblock \showarticletitle{Twitter Sentiment Analysis via Bi-sense Emoji
  Embedding and Attention-based LSTM}. In \bibinfo{booktitle}{\emph{2018 ACM on
  Multimedia Conference}}. ACM, \bibinfo{pages}{117--125}.
\newblock


\bibitem[\protect\citeauthoryear{Clevert, Unterthiner, and Hochreiter}{Clevert
  et~al\mbox{.}}{2015}]%
        {clevert2015fast}
\bibfield{author}{\bibinfo{person}{Djork-Arn{\'e} Clevert},
  \bibinfo{person}{Thomas Unterthiner}, {and} \bibinfo{person}{Sepp
  Hochreiter}.} \bibinfo{year}{2015}\natexlab{}.
\newblock \showarticletitle{Fast and accurate deep network learning by
  exponential linear units (elus)}.
\newblock \bibinfo{journal}{\emph{arXiv preprint arXiv:1511.07289}}
  (\bibinfo{year}{2015}).
\newblock


\bibitem[\protect\citeauthoryear{Eyben, Schuller, Reiter, and Rigoll}{Eyben
  et~al\mbox{.}}{2007}]%
        {eyben2007wearable}
\bibfield{author}{\bibinfo{person}{Florian Eyben}, \bibinfo{person}{Bjorn
  Schuller}, \bibinfo{person}{Stephan Reiter}, {and} \bibinfo{person}{Gerhard
  Rigoll}.} \bibinfo{year}{2007}\natexlab{}.
\newblock \showarticletitle{Wearable assistance for the ballroom-dance
  hobbyist-holistic rhythm analysis and dance-style classification}. In
  \bibinfo{booktitle}{\emph{2007 IEEE International Conference on Multimedia
  and Expo}}. IEEE, \bibinfo{pages}{92--95}.
\newblock


\bibitem[\protect\citeauthoryear{Eyben, W{\"o}llmer, and Schuller}{Eyben
  et~al\mbox{.}}{2010}]%
        {eyben2010opensmile}
\bibfield{author}{\bibinfo{person}{Florian Eyben}, \bibinfo{person}{Martin
  W{\"o}llmer}, {and} \bibinfo{person}{Bj{\"o}rn Schuller}.}
  \bibinfo{year}{2010}\natexlab{}.
\newblock \showarticletitle{Opensmile: the munich versatile and fast
  open-source audio feature extractor}. In
  \bibinfo{booktitle}{\emph{Proceedings of the 18th ACM international
  conference on Multimedia}}. ACM, \bibinfo{pages}{1459--1462}.
\newblock


\bibitem[\protect\citeauthoryear{Gorham}{Gorham}{1988}]%
        {gorham1988relationship}
\bibfield{author}{\bibinfo{person}{Joan Gorham}.}
  \bibinfo{year}{1988}\natexlab{}.
\newblock \showarticletitle{The relationship between verbal teacher immediacy
  behaviors and student learning}.
\newblock \bibinfo{journal}{\emph{Communication education}}
  \bibinfo{volume}{37}, \bibinfo{number}{1} (\bibinfo{year}{1988}),
  \bibinfo{pages}{40--53}.
\newblock


\bibitem[\protect\citeauthoryear{Heimlich and Norland}{Heimlich and
  Norland}{1994}]%
        {heimlich1994developing}
\bibfield{author}{\bibinfo{person}{Joe~E Heimlich} {and}
  \bibinfo{person}{Emmalou Norland}.} \bibinfo{year}{1994}\natexlab{}.
\newblock \bibinfo{booktitle}{\emph{Developing Teaching Style in Adult
  Education. The Jossey-Bass Higher and Adult Education Series.}}
\newblock \bibinfo{publisher}{ERIC}.
\newblock


\bibitem[\protect\citeauthoryear{Heimlich and Norland}{Heimlich and
  Norland}{2002}]%
        {heimlich2002teaching}
\bibfield{author}{\bibinfo{person}{Joe~E Heimlich} {and}
  \bibinfo{person}{Emmalou Norland}.} \bibinfo{year}{2002}\natexlab{}.
\newblock \showarticletitle{Teaching style: where are we now?}
\newblock \bibinfo{journal}{\emph{New directions for adult and continuing
  education}} \bibinfo{volume}{2002}, \bibinfo{number}{93}
  (\bibinfo{year}{2002}), \bibinfo{pages}{17--26}.
\newblock


\bibitem[\protect\citeauthoryear{Kingma and Ba}{Kingma and Ba}{2014}]%
        {kingma2014adam}
\bibfield{author}{\bibinfo{person}{Diederik~P Kingma} {and}
  \bibinfo{person}{Jimmy Ba}.} \bibinfo{year}{2014}\natexlab{}.
\newblock \showarticletitle{Adam: A method for stochastic optimization}.
\newblock \bibinfo{journal}{\emph{arXiv preprint arXiv:1412.6980}}
  (\bibinfo{year}{2014}).
\newblock


\bibitem[\protect\citeauthoryear{Kwon, Chan, Hao, and Lee}{Kwon
  et~al\mbox{.}}{2003}]%
        {kwon2003emotion}
\bibfield{author}{\bibinfo{person}{Oh-Wook Kwon}, \bibinfo{person}{Kwokleung
  Chan}, \bibinfo{person}{Jiucang Hao}, {and} \bibinfo{person}{Te-Won Lee}.}
  \bibinfo{year}{2003}\natexlab{}.
\newblock \showarticletitle{Emotion recognition by speech signals}. In
  \bibinfo{booktitle}{\emph{Eighth European Conference on Speech Communication
  and Technology}}.
\newblock


\bibitem[\protect\citeauthoryear{Li, Zhou, Song, Ran, and Fu}{Li
  et~al\mbox{.}}{2005}]%
        {li2005reliability}
\bibfield{author}{\bibinfo{person}{Xiaoming Li}, \bibinfo{person}{Haotian
  Zhou}, \bibinfo{person}{Shengzun Song}, \bibinfo{person}{Tian Ran}, {and}
  \bibinfo{person}{Xiaolan Fu}.} \bibinfo{year}{2005}\natexlab{}.
\newblock \showarticletitle{The reliability and validity of the Chinese version
  of abbreviated PAD emotion scales}. In
  \bibinfo{booktitle}{\emph{International Conference on Affective Computing and
  Intelligent Interaction}}. Springer, \bibinfo{pages}{513--518}.
\newblock


\bibitem[\protect\citeauthoryear{Li and Sun}{Li and Sun}{2009}]%
        {li2009punctuation}
\bibfield{author}{\bibinfo{person}{Zhongguo Li} {and} \bibinfo{person}{Maosong
  Sun}.} \bibinfo{year}{2009}\natexlab{}.
\newblock \showarticletitle{Punctuation as implicit annotations for Chinese
  word segmentation}.
\newblock \bibinfo{journal}{\emph{Computational Linguistics}}
  \bibinfo{volume}{35}, \bibinfo{number}{4} (\bibinfo{year}{2009}),
  \bibinfo{pages}{505--512}.
\newblock


\bibitem[\protect\citeauthoryear{Mehrabian}{Mehrabian}{1980}]%
        {mehrabian1980basic}
\bibfield{author}{\bibinfo{person}{Albert Mehrabian}.}
  \bibinfo{year}{1980}\natexlab{}.
\newblock \showarticletitle{Basic dimensions for a general psychological theory
  implications for personality, social, environmental, and developmental
  studies}.
\newblock  (\bibinfo{year}{1980}).
\newblock


\bibitem[\protect\citeauthoryear{Mehrabian}{Mehrabian}{1996}]%
        {mehrabian1996pleasure}
\bibfield{author}{\bibinfo{person}{Albert Mehrabian}.}
  \bibinfo{year}{1996}\natexlab{}.
\newblock \showarticletitle{Pleasure-arousal-dominance: A general framework for
  describing and measuring individual differences in temperament}.
\newblock \bibinfo{journal}{\emph{Current Psychology}} \bibinfo{volume}{14},
  \bibinfo{number}{4} (\bibinfo{year}{1996}), \bibinfo{pages}{261--292}.
\newblock


\bibitem[\protect\citeauthoryear{Mohammadi and Vinciarelli}{Mohammadi and
  Vinciarelli}{2011}]%
        {mohammadi2011humans}
\bibfield{author}{\bibinfo{person}{Gelareh Mohammadi} {and}
  \bibinfo{person}{Alessandro Vinciarelli}.} \bibinfo{year}{2011}\natexlab{}.
\newblock \showarticletitle{Humans as feature extractors: combining prosody and
  personality perception for improved speaking style recognition}. In
  \bibinfo{booktitle}{\emph{Systems, Man, and Cybernetics (SMC), 2011 IEEE
  International Conference on}}. IEEE, \bibinfo{pages}{363--366}.
\newblock


\bibitem[\protect\citeauthoryear{Schuller, Steidl, Batliner, Burkhardt,
  Devillers, M{\"u}ller, and Narayanan}{Schuller et~al\mbox{.}}{2010}]%
        {schuller2010interspeech}
\bibfield{author}{\bibinfo{person}{Bj{\"o}rn Schuller}, \bibinfo{person}{Stefan
  Steidl}, \bibinfo{person}{Anton Batliner}, \bibinfo{person}{Felix Burkhardt},
  \bibinfo{person}{Laurence Devillers}, \bibinfo{person}{Christian M{\"u}ller},
  {and} \bibinfo{person}{Shrikanth Narayanan}.}
  \bibinfo{year}{2010}\natexlab{}.
\newblock \showarticletitle{The INTERSPEECH 2010 paralinguistic challenge}. In
  \bibinfo{booktitle}{\emph{Proc. INTERSPEECH 2010, Makuhari, Japan}}.
  \bibinfo{pages}{2794--2797}.
\newblock


\bibitem[\protect\citeauthoryear{Shen, Shepherd, Cui, and Tan}{Shen
  et~al\mbox{.}}{2009}]%
        {ShenSCT09}
\bibfield{author}{\bibinfo{person}{Jialie Shen}, \bibinfo{person}{John
  Shepherd}, \bibinfo{person}{Bin Cui}, {and} \bibinfo{person}{Kian{-}Lee
  Tan}.} \bibinfo{year}{2009}\natexlab{}.
\newblock \showarticletitle{A novel framework for efficient automated singer
  identification in large music databases}.
\newblock \bibinfo{journal}{\emph{{ACM} Trans. Inf. Syst.}}
  \bibinfo{volume}{27}, \bibinfo{number}{3} (\bibinfo{year}{2009}),
  \bibinfo{pages}{18:1--18:31}.
\newblock


\bibitem[\protect\citeauthoryear{Smiljani{\'c} and Bradlow}{Smiljani{\'c} and
  Bradlow}{2009}]%
        {smiljanic2009speaking}
\bibfield{author}{\bibinfo{person}{Rajka Smiljani{\'c}} {and}
  \bibinfo{person}{Ann~R Bradlow}.} \bibinfo{year}{2009}\natexlab{}.
\newblock \showarticletitle{Speaking and hearing clearly: Talker and listener
  factors in speaking style changes}.
\newblock \bibinfo{journal}{\emph{Language and linguistics compass}}
  \bibinfo{volume}{3}, \bibinfo{number}{1} (\bibinfo{year}{2009}),
  \bibinfo{pages}{236--264}.
\newblock


\bibitem[\protect\citeauthoryear{Srivastava, Hinton, Krizhevsky, Sutskever, and
  Salakhutdinov}{Srivastava et~al\mbox{.}}{2014}]%
        {srivastava2014dropout}
\bibfield{author}{\bibinfo{person}{Nitish Srivastava},
  \bibinfo{person}{Geoffrey Hinton}, \bibinfo{person}{Alex Krizhevsky},
  \bibinfo{person}{Ilya Sutskever}, {and} \bibinfo{person}{Ruslan
  Salakhutdinov}.} \bibinfo{year}{2014}\natexlab{}.
\newblock \showarticletitle{Dropout: a simple way to prevent neural networks
  from overfitting}.
\newblock \bibinfo{journal}{\emph{The journal of machine learning research}}
  \bibinfo{volume}{15}, \bibinfo{number}{1} (\bibinfo{year}{2014}),
  \bibinfo{pages}{1929--1958}.
\newblock


\bibitem[\protect\citeauthoryear{Sun, Chen, Zhang, Guo, and Liu}{Sun
  et~al\mbox{.}}{2016}]%
        {sun2016thulac}
\bibfield{author}{\bibinfo{person}{Maosong Sun}, \bibinfo{person}{Xinxiong
  Chen}, \bibinfo{person}{Kaixu Zhang}, \bibinfo{person}{Zhipeng Guo}, {and}
  \bibinfo{person}{Zhiyuan Liu}.} \bibinfo{year}{2016}\natexlab{}.
\newblock \bibinfo{booktitle}{\emph{Thulac: An efficient lexical analyzer for
  chinese}}.
\newblock \bibinfo{type}{{T}echnical {R}eport}. \bibinfo{institution}{Technical
  Report}.
\newblock


\bibitem[\protect\citeauthoryear{Thammasan, Fukui, and Numao}{Thammasan
  et~al\mbox{.}}{2017}]%
        {thammasan2017multimodal}
\bibfield{author}{\bibinfo{person}{Nattapong Thammasan},
  \bibinfo{person}{Ken-ichi Fukui}, {and} \bibinfo{person}{Masayuki Numao}.}
  \bibinfo{year}{2017}\natexlab{}.
\newblock \showarticletitle{Multimodal Fusion of EEG and Musical Features in
  Music-Emotion Recognition.}. In \bibinfo{booktitle}{\emph{AAAI}}.
  \bibinfo{pages}{4991--4992}.
\newblock


\bibitem[\protect\citeauthoryear{Trendowicz and Jeffery}{Trendowicz and
  Jeffery}{2014}]%
        {Trendowicz2014Classification}
\bibfield{author}{\bibinfo{person}{Adam Trendowicz} {and} \bibinfo{person}{Ross
  Jeffery}.} \bibinfo{year}{2014}\natexlab{}.
\newblock \bibinfo{booktitle}{\emph{Classification and Regression Trees}}.
\newblock \bibinfo{publisher}{Springer International Publishing}. 1174–1176
  pages.
\newblock


\bibitem[\protect\citeauthoryear{Wurtzel and Dominick}{Wurtzel and
  Dominick}{1971}]%
        {wurtzel1971evaluation}
\bibfield{author}{\bibinfo{person}{Alan~H Wurtzel} {and}
  \bibinfo{person}{Joseph~R Dominick}.} \bibinfo{year}{1971}\natexlab{}.
\newblock \showarticletitle{Evaluation of television drama: Interaction of
  acting styles and shot selection}.
\newblock \bibinfo{journal}{\emph{Journal of Broadcasting \& Electronic Media}}
  \bibinfo{volume}{16}, \bibinfo{number}{1} (\bibinfo{year}{1971}),
  \bibinfo{pages}{103--110}.
\newblock


\bibitem[\protect\citeauthoryear{Zhang, Provost, and Essi}{Zhang
  et~al\mbox{.}}{2016}]%
        {zhang2016cross}
\bibfield{author}{\bibinfo{person}{Biqiao Zhang}, \bibinfo{person}{Emily~Mower
  Provost}, {and} \bibinfo{person}{Georg Essi}.}
  \bibinfo{year}{2016}\natexlab{}.
\newblock \showarticletitle{Cross-corpus acoustic emotion recognition from
  singing and speaking: A multi-task learning approach}. In
  \bibinfo{booktitle}{\emph{Acoustics, Speech and Signal Processing (ICASSP),
  2016 IEEE International Conference on}}. IEEE, \bibinfo{pages}{5805--5809}.
\newblock


\bibitem[\protect\citeauthoryear{Zhou, Jia, Wang, Dong, Yin, and Lei}{Zhou
  et~al\mbox{.}}{2018}]%
        {zhou2018inferring}
\bibfield{author}{\bibinfo{person}{Suping Zhou}, \bibinfo{person}{Jia Jia},
  \bibinfo{person}{Qi Wang}, \bibinfo{person}{Yufei Dong},
  \bibinfo{person}{Yufeng Yin}, {and} \bibinfo{person}{Kehua Lei}.}
  \bibinfo{year}{2018}\natexlab{}.
\newblock \showarticletitle{Inferring Emotion from Conversational Voice Data: A
  Semi-supervised Multi-path Generative Neural Network Approach}.
\newblock  (\bibinfo{year}{2018}).
\newblock


\end{thebibliography}

\end{document}